# Weak Mixing Matrix Structures in the Broken Mirror Symmetry Model

Igor T. Dyatlov *
**Scientific Research Center "Kurchatov Institute"**
**Petersburg Institute of Nuclear Physics, Gatchina, Russia**

A model of symmetry violation is described for a system that can spontaneously choose the left-handed or right-handed character of weak currents. For the hierarchic structure of fermion mass spectra, such a mirror symmetry system allows reproduction of all qualitative properties of weak mixing matrices for both quarks (CKM matrix) and leptons (PMNS matrix),and at that, without additional numerical fitting of model parameters. The hierarchy of CKM matrix elements is directly stipulated by the hierarchical mass spectrum of quark generations. The qualitative properties of the PMNS matrix arise for the inverse character of the neutrino spectrum ($m_3$ is the smallest mass) and Dirac nature of SM neutrino. The comparatively small value of the neutrino mixing angle $\theta_{13}$ is a consequence of the smallness of $m_3$ and the smallness of the charged lepton mass ratio $m_e/m_\mu$.

## 1. Introduction

For many years, it was believed that the character of the coordinate system (left ($L$) or right ($R$)) used to study a given phenomenon could not be defined by physical methods. Spatial parity non-conservation has presented such an opportunity. By stating that upon radioactive decay, the majority of electrons move in the direction of magnetic field, one determines the direction of the axial vector of magnetic field strength and thereby the character of the coordinate system in which a given phenomenon is being considered.

Unlike choosing the origin of the coordinate system and the direction of its axes in space, setting the coordinate system to be either $L$- or $R$-handed by physical methods is possible owing to parity non-conservation. Such a situation seemed unacceptable yet to the pioneers Lee and Young [1]. They promptly proposed [2] a model of a system that eliminated the paradox, supplementing the observed system of particles (in modern terms, the Standard Model (SM)) with an identical system of heavy, and therefore yet undiscovered, analogs with opposite weak properties ($L \leftrightarrow R$). Later, these analogs were termed "mirror particles".

Experimental attempts to detect mirror particles and various theoretical suggestions around Lee and Young's idea have continued to the present time (see reviews [3,4]). Numerous studies and a large variety of proposed approaches indicate that direct parity non-conservation has not seemed satisfactory despite the success and consistency of the SM, which incorporates this non-conservation in its immediate structure.

*E-mail: dyatlov@thd.pnpi.spb.ru

All ongoing research into mirror particles mostly focuses on experimental possibilities to detect and identify such particles and their influence on SM effects [7,8]. Almost no papers exist which try to describe the observed properties of SM particles by means of mirror models. This paper is an attempt to show that the qualitative structure of weak mixing matrices (WMM) for both quarks and leptons can be explained by the existence of very heavy mirror particles. At that, the qualitative properties of the WMM are based exclusively on the qualitative properties of the SM particle system, that is, the hierarchical order of the fermion mass spectrum and the violated weak symmetry group $SU(2)$ that forms the basis of weak interactions in the SM.

Hierarchies of fermion mass generations are of crucial importance. They define the hierarchical order of quark WMM non-diagonal elements and may be instrumental in producing a different lepton WMM form. At that, no additional fitting of the model's constants is required. The exact qualitative match with the observed WMM form does not depend on the number of parameters used, their values and complexity.

Many authors have tried to link the hierarchical character of fermion mass spectra with the quark WMM [9] and to invent the dynamics of its generation (see [10] and references therein). In this paper, assuming the existence of the hierarchical spectra, we confirm their fundamental role in the creation of specific qualitative forms of quark and lepton WMMs. At the same time, mirror fermions themselves must be much heavier than SM fermions, which is one of the main conditions for the appearance of correct structures—light SM particle masses and their WMMs.

We believe that this very system could correspond to an experimental absence of mirror particles and their extraordinarily insignificant influence on SM effects and physics beyond the SM.

As noted in [11], the representation of a mass matrix as a sum of separable matrices with an hierarchy of addend values not only results in the hierarchical mass spectrum but also permits an easy reproduction of quark WMM qualitative properties (CKM matrices) as a result of just the weak $SU(2)$ symmetry, with no fitting of parameters required.

Separability means that the transition between the system's initial and final states occurs through some intermediate state. The only difference of this intermediate state from the initial and final states is the mass, while all other quantum numbers are preserved. This is exactly what characterizes mirror particles, since the direct scalar transition of a fermion into another fermion means $R \leftrightarrow L$. Moreover, intermediate states must be much heavier than the initial and final state masses, so that momenta in propagators describing the process (Fig.1) can be neglected.

Transitions (Fig.1a) from SM particles to heavy mirror states are defined by the mass parameters $A, B$ of fermions prior to mirror symmetry (MS) violation (Eq.(3)). As a result, the mirror mechanism appears to be less complicated than the transitions involving Yukawa couplings and vacuum averages of scalar bosons [10] (Fig.1b).

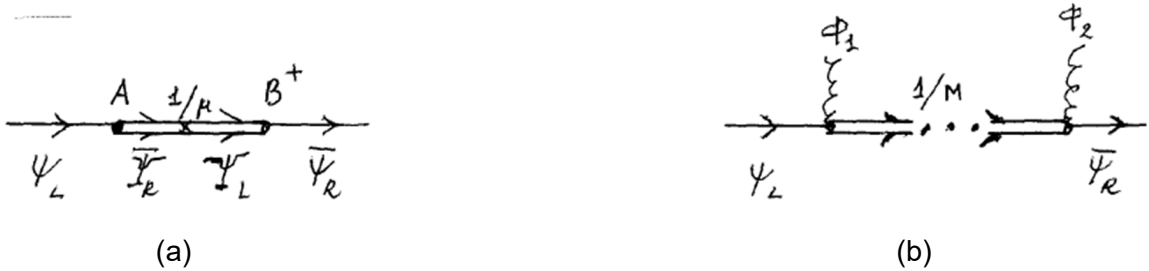

(a)                                                                                    (b)

Fig.1: Fermion mass formation mechanism in this paper (a) and in [10,20] (b); $\Phi_1 \Phi_2 -$ "Higgs vacuum averages".

Formulae for light quark and lepton masses (corresponding to SM states) are then governed by similar expressions to the see-saw mechanism [12]. These masses arise upon spontaneous violation of the initial MS. Heavy masses of mirror particles play here the role of large see-saw Majorana masses. Dirac masses are replaced by the parameters $A, B$ of fermion masses in the initial unviolated MS-Lagrangian.

The above hypothesis has been developed by the author in a number of papers [13]. Applying it to quarks and charged leptons requires consideration of only qualitative properties; that is, very large masses of mirror particles and the weak $SU(2)$-symmetry of the full MS model. The erroneous premise in [13] that light Dirac neutrinos in the MS models could be built even for the conventional see-saw mechanism (with heavy Majorana masses) leads to an incorrect conclusion (the third reference in [13] - error correction). But then light neutrinos are also of the Majorana type. In such a scenario, it appears impossible to reproduce the observed properties of the lepton WMM (PMNS matrix) for any character of the neutrino spectrum without introducing additional, unsubstantiated conditions for constants relating light and heavy particles. Moreover, with Majorana neutrinos in the MS-model, it is impossible to find any convincing qualitative mechanism for the appearance of the inconspicuous properties of the PMNS matrix.

The MS violation model described in this paper interprets the neutrino system by exact analogy with the quark and charged lepton system; i.e., neutrinos are also considered to be Dirac ones. The MS-Lagrangian then lacks Majorana terms. Neutrino masses become small if their mirror counterparts are very heavy. The spectrum must be inverse and neutrinos are only of the Dirac character. Such a system is able to qualitatively reproduce both small neutrino masses and the observed structure of the PMNS matrix[1] without any additional fitting of the model's constants. A relatively small sinus of the neutrino flavour mixing angle $sin\,\theta_{13}$, the only conspicuous feature of the PMNS matrix [14], becomes in this case a natural and easy to explain feature.

The primary purpose of this paper is the derivation of expressions for both WMMs. Other problems related with the use of MS-mechanisms are discussed in the author's previous papers and are briefly summarized under Conclusions in this paper. Section 2 discusses the general scheme for MS introduction. Section 3 describes the approximate scheme of MS violation constructed by exact analogy with the weak symmetry violation in the SM. In Section 4, a mass matrix for the quark system is obtained, as well as its corresponding WMM. Section 5 examines conditions for the appearance in the scenario under consideration of a lepton WMM with qualities close to the observed matrix. Appendix 1 presents the general derivation of a mass matrix for the models and their possible WMMs. Appendix 2 proposes a hypothesis of a reason for the appearance of spectra with fermion mass hierarchies which is not dependent on the properties of interactions.

## 2. Mirror Symmetry

In [11,13], the MS concept implies the full identity of $R$- and $L$-systems before symmetry violation. This approach differs from other mirror systems proposed in literature [3-7], where $R$- and $L$-systems may alternatively incorporate different interactions, different representations of the various groups, or different $R$-, $L$-vector bosons and even photons. In our opinion, identical systems most closely correspond to the original idea of Lee and Young [2]. The model proposed here is based on the difference in weak properties $(R \leftrightarrow L)$ appearing only after symmetry violation.

In the MS-system, the Lagrangian only depends on the full Dirac operators:

$$\Psi_{LR} = \psi_L + \Psi_R\,(T_W = 1/2)\,, \quad \Psi_{RL} = \psi_R + \Psi_L\,(T_W = 0), \tag{1}$$

---

- doublets and singlets of the weak isospin $T_W$. In Eq.(1), we omitted all other quantum numbers: $\bar{u}$ (up) and $\bar{d}$ (down) flavors, generation indices $n, n' = 1,2,3$, etc. Eq.(1) represents massive Dirac fermions—quarks and leptons.

The term "mirror transformation" used here for an identical operation that should obviously preserve the invariance of the Lagrangian:

$$R \leftrightarrow L\,, \qquad \psi \leftrightarrow \Psi\,. \tag{2}$$

The kinetic part and any gauge interaction of the SM can be written out using the Eq.(1) operators and are automatically, just owing to the chiral properties, subdivided into $\Psi$- and $\psi$-components. Weak interaction is produced by the vector current of doublets $\Psi_{LR}$ and is also a sum of the currents $\Psi_R$ and $\psi_L$. MS-interactions with scalars defining mass spectrum properties are discussed in Section 3, but the masses of states (1) themselves are directly included in the MS-Lagrangian:

$$\mathcal{L}_0 \; = \mathcal{L}_{SM}(\Psi_{LR}, \Psi_{RL}) + A\bar{\Psi}_{LR}\Psi_{LR} + B\bar{\Psi}_{RL}\Psi_{RL}\,, \tag{3}$$

where $A$ and $B$ are, in general terms, 3x3 matrices over generation indices. These terms are $SU(2)$-invariant, which means:

$$A^{(\bar{u})} \; \equiv \; A^{(\bar{d})}, \quad B^{(\bar{u})} \; \neq \; B^{(\bar{d})}\,. \tag{4}$$

Owing to its chiral properties, terms (3) with the masses $A, B$ are the only ones that relate the $\psi$- and $\Psi$-components with each other (see Eq.(18)). Apparently, the matrices $A$ and $B$ must be Hermitian. The mechanism, violating the symmetry (2), will be built here by exact analogy with the SM scenario.

The task of introducing the $\psi$ and $\Psi$ asymmetry, while keeping the other properties fully identical, can be solved only if the system has two populations of states that differ only in the $R$- and $L$-properties. This corresponds to two possible ground states (vacuums) with different weak properties, which means that upon breaking of the $\psi \leftrightarrow \Psi$ symmetry, the weak current in one of the states, $|L\rangle$, becomes left-handed for light particles $\psi$ and right-handed for heavy $\Psi$. In the other state, $|R\rangle$, the properties of weak currents are opposite.

Two types of vacuum expectations can appear in the system under consideration:

$$\langle R|\varphi_1|R \rangle \;\equiv\; \langle L|\varphi_1|L \rangle \,, \tag{5}$$

where the equality takes place because of the assumption that all other properties are identical and

$$\left| \langle R|\varphi_2|L \rangle \right| \;=\; \left| \langle L|\varphi_2^+|R \rangle \right| . \tag{6}$$

Apparently, the field $\varphi_1$ is a scalar and $\varphi_2$ is a pseudoscalar. It is necessary to have at least two types of "Higgs" scalars (pseudo-) and the corresponding Yukawa couplings giving rise to $\psi$ and $\Psi$ masses. The difference in these masses must be due to the existence of different minima of the scalar field potential $V(\varphi_1 \varphi_2)$.

Before proceeding to the actual construction of a suitable model, we should discuss some important aspects (positive and negative) of the chosen standard method of spontaneous violation using scalar vacuum expectations.

The problem of the strong Yukawa coupling becomes critical here: very heavy fermions are an integral part of the overall system. Of course, we cannot solve this problem. Moreover, only heavy mirror $\Psi$ particles interact with the observed scalar $H$ in the Lagrangian itself (see Eq.(10)). We will show, however, that for "light" Dirac fermions of the SM, the mechanism of mass formation (section 3) preserves the usual perturbative (at the existing masses and vacuum average $\eta \approx 246$ GeV) interaction with the Higgs particle $H$ (coupling constant $h \sim m/\eta$). This property is a direct consequence of the system being invariant relative to the weak isospin $SU(2)$ and does not require any additional assumptions.

It appears as though, by creating the SM, the "nature" (in our model) isolated the perturbative part of a large general system.[2] All coupling constants remain here in the perturbative part. For this purpose, SM particles had to be broken away from mirror states so that the two parts would have no strong relationship. This can be achieved through two important observed features of the SM.

1. Although the full system – SM particles + mirror generations ($L, R$-symmetrical system) – does not have chiral anomalies [17], its low-energy part (SM), as well known, per se lacks anomalies and can therefore be renormalized and perturbative. A complex cancellation takes place for quark and lepton anomalous contributions. An absence of low-energy cancellation would indicate a strong coupling between all levels of the system, in which case no meaningful breaking of low and high states could happen.

---

[2] Strong quark interactions do not play any role in the formation of fundamental masses. Their impact is insignificant at very small distances.

2. The smallness of the Higgs boson mass ($m_H \approx 126$ GeV [15]) not only preserves perturbative unitarity in processes involving longitudinal vector bosons ($m_H < 1$ TeV), but also ensures that scalar self-interaction $\lambda$ ($V = \phi^4$, $m_H < 0.5$ TeV) is perturbative. Large masses of the $H$ boson would also lead to a strong coupling with high-energy states.

To conclude this section, it should be noted that the selected MS system offers an answer to the question that remains unanswered in other scenarios [3]—is it true that all mirror particles are heavier than SM particles? The proposed mechanism reproduces the WMM only if the process of SM mass matrix formation involves very heavy mirror fermions with masses much larger than SM masses.

Yet, the origin of the Dirac fermion mass hierarchy in the SM itself continues to remain unexplained (see [10] and Appendix 2).

## 3. Mirror Symmetry Spontaneous Violation Model

Based in the section 2 conclusions, let us consider the Yukawa couplings of mirror symmetry operators (1) with complex isodoublets $\varphi_1$ (scalar) and $\varphi_2$ (pseudoscalar). By exact analogy with the SM:

$$
\begin{aligned}
\mathcal{L}' \;=\; & (h_1^{(\bar{u})})^{n'}_n \left( \bar{\Psi}^n_{LR}, \varphi_1^c \right) \Psi^{(\bar{u})}_{RL\,n'} + (h_2^{(\bar{u})})^{n'}_n \left( \bar{\Psi}^n_{LR}\gamma_5, \varphi_2^c \right) \Psi^{(\bar{u})}_{RL\,n'} + c.c. \;+ \\
& + \left( h_1^{(\bar{d})} \right)^{n'}_n \left( \bar{\Psi}^n_{LR}, \varphi_1 \right) \Psi^{(\bar{d})}_{RL\,n'} + \left( h_2^{(\bar{d})} \right)^{n'}_n \left( \bar{\Psi}^n_{LR}\gamma_5, \varphi_2 \right) \Psi^{(\bar{d})}_{RL\,n'} + c.c.
\end{aligned}
\tag{7}
$$

Eq.(7) writes out all state indices describing the system under consideration: the flavor $f = \bar{u}, \bar{d}$, generation indices $n, n' = 1,2,3$. The round brackets $(\bar{\Psi}, \varphi)$ denote products of isodoublets, $\varphi^c = i\sigma_y \varphi^+$. Further on, to simplify our formulae, most indices will be omitted. The $h$ matrices can be arbitrary.

Substituting operators (1) in Eq.(7), we obtain:

$$
\begin{aligned}
\mathcal{L}' \;=\; & h_1\Big[ (\bar{\Psi}_R, \varphi_1)\Psi_L + (\bar{\psi}_L\varphi_1)\psi_R \Big] + h_2\Big[ -(\bar{\Psi}_R, \varphi_2)\Psi_L + (\bar{\psi}_L, \varphi_2)\psi_R \Big] + \cdots \;= \\
\;=\; & \left( \bar{\Psi}_R, [h_1\varphi_1 - h_2\varphi_2] \right)\Psi_L + \left( \bar{\psi}_L, [h_1\varphi_1 + h_2\varphi_2] \right)\psi_R + \cdots
\end{aligned}
\tag{8}
$$

Identity of the $\Psi$ and $\psi$ systems after MS violation requires that the $h_i$ matrices be equal:

$$h_1 \equiv h_2 = h. \tag{9}$$

In this case, light particle "worlds" arising upon various types of MS spontaneous violation are in fact (see Eq.(12)) absolutely identical (with the exception of the $R \leftrightarrow L$ weak properties).Then the Lagrangian (8) takes the following form:

$$\mathcal{L}' = h(\bar{\Psi}_R, \Phi_1)\Psi_L + h(\bar{\psi}_L, \Phi_2)\psi_R + \cdots$$
$$\Phi_1 = \varphi_1 - \varphi_2, \quad \Phi_2 = \varphi_1 + \varphi_2. \tag{10}$$

In the MS world, the $\Phi_1$ and $\Phi_2$ operators do not have certain parity. In a violated MS, however, this feature does not change the properties of the system. As is the case with the SM, the same bosons $\Phi_1$ and $\Phi_2$ produce a Yukawa coupling for both quarks and leptons. This is necessary to prevent an increase in the number of Goldstone phases in boson $SU(2)$ doublets. Three phases of one of the $\Phi_1$-, $\Phi_2$-operators achieving a vacuum expectations are sufficient to make all three weak vector $W_\mu$ bosons heavy.

Let us take as a potential for the scalars $V(\Phi_1, \Phi_2)$ - a $\Phi_1\Phi_2$ symmetrical expression built by exact analogy with the SM:

$$V(\Phi_1, \Phi_2) = \kappa|\Phi_1|^2|\Phi_2|^2 - \frac{\rho^2}{2}\Big(|\Phi_1|^2 + |\Phi_2|^2\Big) + \frac{\lambda}{4}\Big(|\Phi_1|^4 + |\Phi_2|^4\Big). \tag{11}$$

At large $\kappa$, the deepest minima of $V$ split the system both by $\Phi_1$ and $\Phi_2$ and by $\Psi$ and $\varphi$. For vacuum averages, the solution of Eq.(11) is as follows:

$$\langle \Phi_2 \rangle = 0, \quad \langle \Phi_1 \rangle^2 = \frac{\rho^2}{\lambda} = \eta^2,$$
$$\langle \Phi_1 \rangle = 0, \quad \langle \Phi_2 \rangle^2 = \frac{\rho^2}{\lambda}. \tag{12a}$$

Of interest to the SM is variant (12a), where the operator $\Phi_1$ represents a system with one neutral scalar $H$ (the Higgs boson of the SM) and Goldstone phases serving to create non-zero $W$-boson masses. At that, the tree-level mass of the other boson $\Phi_2$, equal to

$$M_{\Phi_2}^2 = \kappa\frac{\rho^2}{\lambda} - \frac{\rho^2}{2}, \tag{13}$$

can be made as heavy as needed at large $\kappa$:

$$\kappa \eta^2 \;\gg\; \frac{\rho^2}{2}\,. \tag{14}$$

The operator $\Phi_2$ creates four scalar states: two charged ($\pm$) and two neutral, as it happens in the case of $K, \overline{K}$-mesons.

We have to discuss now an unfortunate circumstance associated with the unavoidable presence of non-perturbative Yukawa interaction in the scenario under consideration. If $\Phi_1$ is identified with the Higgs scalar $H$, then the quantity $\langle \Phi_1 \rangle = \eta$ is defined by the mass of the $W$-boson ($M_W = \frac{g_W \eta}{2}, \eta \simeq 246$ GeV). (Note that even in the presence of strong interactions, $W$ masses, when produced spontaneously, are mainly defined by the pole Goldstone contribution. Other contributions to the mass operator represent weak corrections proportional to the factor $(k^2 g_W - k_\mu k_\nu)$ that have no pole at $k^2 = 0$.) Heavy $\Psi$ masses therefore mean large values of $h \simeq M_\Psi / \eta \gg 1$. This practically suspends any further quantitative use of the proposed scheme (see [18]).

At the same time, the interaction of the standard Higgs boson $H$ with SM light fermions $\psi$ is consistent with the perturbative treatment we had in the SM. In fact, although the Higgs scalar $\Phi_1$ in Eq.(10) directly interacts only with the mirror states $\Psi$, diagonalization of the mass matrices of the full system $\psi, \Psi$ results in eigenfunctions of $\Psi_{M^-}$ and $\psi_{\mu^-}$ type massive states (see Section 7 in [11]):

$$\begin{aligned}
\Psi_M &= \sum \left[ \Psi + O\left( \left( \frac{\mu_\psi}{M_\Psi} \right)^{1/2} \right) \psi \right], \\
\psi_\mu &= \sum \left[ \psi + O\left( \left( \frac{\mu_\psi}{M_\Psi} \right)^{1/2} \right) \Psi \right].
\end{aligned} \tag{15}$$

where $\mu_\psi$ are quantities in the order of SM fermion masses $\mu_\psi \ll M_\Psi$. The right side of Eq.(15) represents sums over generation indices. Then the interaction $h_\psi \bar{\psi}_\mu \psi_\mu \Phi_1$ will have a coupling constant typical of the SM. From Eqs.(10) and (15) we have:

$$h_\psi \;=\; h \left( \frac{\mu_\psi}{M_\Psi} \right)_1^{1/2} \left( \frac{\mu_\psi}{M_\Psi} \right)_2^{1/2} \sim \frac{M_\Psi}{\eta} \cdot \frac{\mu_\psi}{M_\Psi} \sim \frac{\mu_\psi}{\eta}\,. \tag{16}$$

This result is a necessary consequence of the weak $SU(2)$-symmetry.

Symmetry in fact means that the pole $q^2$ in the transverse part of the vector boson propagator

$$\Delta_{\mu\nu} \ = \ \frac{g_{\mu\nu} - (q_\mu q_\nu)/q^2}{M_W^2 - q^2}$$

must, in the invariant gauge, be cancelled out with Goldstone contributions formed by the phases of the scalar $\Phi_1$. The cancellation must occur for all contributions in all diagrams of fermions $\psi$ interacting with $W$, which is what actually happens owing to relations (15) and (16).

## 4. Masses and Quark WMM

The Lagrangian (10) and vacuum average (12a) produce Dirac mass terms of mirror particles

$$\mathcal{L}' \ = \ M_\Psi \bar{\Psi}_R \Psi_L + \cdots .$$ (17)

Along with the mass matrices $A$ and $B$ of the MS-states $\Psi_{LR}$ and $\Psi_{RL}$ from the initial Lagrangian (3), which, after mirror symmetry violation, realize transfers $\Psi \leftrightarrow \psi$:

$$\mathcal{L}'_0 \ = \ A\bar{\Psi}_{LR}\Psi_{LR} + B\bar{\Psi}_{RL}\Psi_{RL} \ = \ A\bar{\psi}_L\Psi_R + B\bar{\psi}_R\Psi_L + c.c. \cdots ,$$ (18)

from Eqs.(17) and (18) we can obtain a system of equations for the mass matrices of particles $\psi$.

For this purpose, let us take one of the systems of matrices (in the space of generation indices) - $h_i$ in Eqs.(3),(10), or $A, B$ in (18) - as diagonal. This can always be done without losing the generality of the Lagrangian. The simplest and representative system of equations for masses $\psi$ is obtained for a diagonal form of the matrices $h_i$, i.e., when the matrices $M_\Psi$ in Eq.(17) are diagonal matrices with eigenvalues $M_n$, $n = 0,1,2$ close (at $M \gg |A|$, [13]) to the masses of heavy mirror fermions (Eq.(15)).

The mass matrix of light particles $\psi$ can then be obtained by two methods.

1. The diagram in Fig. 1 permits writing out the expression for the tree-level approximation as a sum of separable matrices

$$\left(\mathcal{M}_{LR}^{(f)}\right)_a^b = \sum_{n=0}^{2} A_a^n \frac{1}{M_n^{(f)}} B_n^{+(f)b}.$$ (19)

Eq.(19) arises when all $|M_n|$ are large, and the momenta $|\hat{p}| \simeq m_\psi \ll M$ can be neglected in the propagators $\Psi$. The tensors $A$ and $B$ are formed by three vectors in the space of generations: $A^n = (A_1^n, A_2^n, A_3^n)$ and same for $B$, $a, b = 1,2,3$.

2. Expressions (17),(18) for each $f = \bar{u}, \bar{d}$ represent the 6th order matrices of the components $\psi, \Psi$

$$
\mathcal{M}_{LR} = \begin{array}{c} \quad \overline{\Psi}_L \qquad \overline{\psi}_L \\ \hline \begin{vmatrix} M & B^+ \\ A & 0 \end{vmatrix} \begin{array}{c} \Psi_R \\ \psi_R \end{array} \end{array}
\tag{20}
$$

with the diagonal matrix $M$.

$\mathcal{M}_{LR}$, Eq.(20), is a direct generalization of the mass matrix of the see-saw mechanism for the case of three fermion generations provided that

$$
|M| \gg |A|, |B|.
\tag{21}
$$

Besides the three large eigenvalues $\mathcal{M}_n \approx M_n$, we have three light masses $\mu_\psi$. To obtain them, let us calculate an inverse matrix to Eq.(20) [19]:

$$
\mathcal{M}_{LR}^{-1} = \begin{vmatrix} 0 & A^{-1} \\ (B^+)^{-1} & -(B^+)^{-1}MA^{-1} \end{vmatrix}.
\tag{22}
$$

In the lowest approximation of (21), for $\mu_\psi$ we also have a sum of separable matrices (19)—a matrix inverse to $(B^+)^{-1}MA^{-1}$. At that, corrections for masses (19) and eigenfunctions (15) in the order of

$$
\left(\frac{A}{M}, \frac{B}{M}\right) \approx \left(\frac{A^2}{M}\frac{1}{M}\right)^{1/2} \approx \left(\frac{B^2}{M}\frac{1}{M}\right)^{1/2} \approx \left(\frac{m_\psi}{M}\right)^{1/2}, \ (A \sim B).
\tag{23}
$$

are neglected. These quantities are assumed to be smaller than the mass ratios of $\mu_\psi$ themselves (ratios of generation mass hierarchy). Corrections in Eq.(19) related with the mass hierarchy $m_i/m_k$ will be taken into account. They define the WMM structure and the magnitude of its small elements, whereas $\mu_\psi/M$ could be assumed to be very small at very large $|M|$.

Diagonalization of matrix (19) in which the smallness of the terms $n = 0,1,2$ obeys a hierarchical order is performed in [11] and largely repeated in Appendix 1. State eigenvalues and eigenfunctions in the space of generations are found. WMM quark coefficients are presented (CKM matrix). Eigenvalues, i.e., masses of the observed generations, form a hierarchical spectrum with consistently decreasing contributions from the terms $n = 0,1,2$.

In the lowest approximation of the mass hierarchy, for each of generations I, II, III (I being the highest order) we obtain the following results:

$$m_I = \frac{|A_0||B_0|}{M_0}, \quad m_{II} = \frac{D_2^{1/2}(|A_0|^2, |A_1|^2)D_2^{1/2}(|B_0|^2, |B_1|^2)}{|A_0||B_0|M_1},$$
$$m_{III} = \frac{D_3^{1/2}(|A_0|^2, |A_1|^2, |A_2|^2)D_2^{1/2}(|B_0|^2, |B_1|^2, |B_2|^2)}{D_2^{1/2}(|A_0|^2, |A_1|^2)D_2^{1/2}(|B_0|^2, |B_1|^2)M_2}.$$

(24)

$A_n$ and $B_n$ are column-vectors forming the matrices $A$ and $B$, $|A_n|$, $|B_n|$ are vector lengths, $D$ are determinants calculated using the formulae in Appendix 1. Eqs.(24) result from formulae (A1.13) – (A1.15).

Apparently, the mass hierarchy of generations $m_I \gg m_{II} \gg m_{III}$ in Eq.(24) can be presented as a hierarchy of the parameters

$$M_0 \ll M_1 \ll M_2.$$

(25)

However, the parameters $A$ and $B$, describing masses before MS violation, can follow the same pattern. Since the hierarchical spectrum of generations is evident in all SM fermions[3], hierarchy may be a common property of fermion spectra (see Appendix 2).

To understand the structure of the quark WMM (the CKM matrix), it is essential that in the lowest approximation of the hierarchy, i.e., in calculations taking into account Eq.(25), eigenfunctions of the left states in matrix (19) are defined only by those vectors $A_0, A_1, A_2$ that are independent of the flavor $f = \bar{u}, \bar{d}$. For generations I, II, III, we obtain the orthonormalized vectors in the space of generations (Eqs. (A1.19) – (A1.21):

$$\phi_I \simeq \frac{1}{|A_0|}A_0, \quad \phi_{II} \simeq \frac{1}{|A_1 - \frac{(A_0^+, A_1)}{|A_1|^2}A_0|}\left(A_1 - \frac{(A_0^+, A_1)}{|A_0|^2}A_0\right),$$
$$\phi_{III} \simeq \frac{1}{|[A_0^+, A_1^+]|}[A_0^+, A_1^+],$$

(26)

---

[3] With the probable exception of neutrino, see Section 5.

where $(A_0^+, A_1)$ and $(A_0^+, A_1^+)$ are scalar and vector products of the vectors; generation I is the heaviest.

The WMM matrix elements are scalar products of the eigenvectors:

$$V_{mn} \equiv \left( \phi_m^{(\bar{u})}, \phi_n^{(\bar{d})} \right). \tag{27}$$

In the lowest approximation of the quark WMM, $V$ appears to be a unitary matrix. Taking subsequent approximations into account results in minor changes to the diagonal terms and the appearance of small, non-zero, non-diagonal terms. In Appendix 1, expressions for the elements of the matrix $V$ are written out also for the arbitrary complex vectors $A, B$ ($|A|, |B| \ll M$) taking necessary approximations into account. To simplify the representation of the hierarchy of the $V$ elements for the flavors $\bar{u}, \bar{d}$, let us neglect the complexities of $A$ and $B$, i.e., the possible CP-violating phases (which has no impact on the hierarchy).

From formulae (A1.22), (A1.23) we obtain, for the main approximation under consideration, the elements of the quark WMM proposed by Wolfenstein [20]:

$$V_{ud} \simeq V_{cs} \simeq V_{tb} \simeq 1 \,, \quad V_{cd} = -V_{us} \,, \quad V_{ts} = -V_{cb} \,. \tag{28}$$

From the formulae for the minor elements of $V$ in Appendix 1 let us isolate the combinations of constants (24) representing the quark masses: $m_I \to m_t$ or $m_b$, $m_{II} \to m_c, m_s$, $m_{III} \to m_u, m_d$, and introduce angles $\beta_{nm}^{\bar{u},\bar{d}}$ between the vectors $B_n^{(f)} = (B_n^1, B_n^2, B_n^3)^f$.

For the elements of the quark WMM, we have (Eqs.(A1.22) and (A1.23)):

$$|V_{cd}| = \left| \frac{m_d}{m_s} f(\beta^{(d)}) - \frac{m_u}{m_c} f(\beta^{(\bar{u})}) \right|, \quad f(\beta) = \frac{\cos \beta_{12} - \cos \beta_{01} \cos \beta_{02}}{\sin \beta_{12} \cos \beta_{012}} \,,$$

$$\sin^2 \beta_{012} = \frac{\cos^2 \beta_{01} + \cos^2 \beta_{02} - 2 \cos \beta_{12} \cos \beta_{01} \cos \beta_{02}}{\sin^2 \beta_{12}} \,.$$

Here $\beta_{012}$ is the angle between $B_0$ and the perpendicular to the plane $B_1, B_2$. For $V_{ts}$, we have:

$$|V_{ts}| = \left| \frac{m_s}{m_b} (\operatorname{ctg} \beta_{01})^{(\bar{d})} - \frac{m_c}{m_t} (\operatorname{ctg} \beta_{01})^{\bar{u}} \right|. \tag{29}$$

For the adopted quark mass ratios [15], the main terms in Eqs.(29) and (30) - the first terms of the sum on the right side of the formulae – are quantities of the order:

$$|V_{cd}| \; \sim \; \frac{m_d}{m_s} \; > \; |V_{ts}| \; \sim \; \frac{m_s}{m_b},\qquad(30)$$

which corresponds to the observed values of the CKM matrix. General expressions for the elements $V_{td}$ and $V_{ub}$ are provided in Appendix 1 (Eqs.(A1.24) and (A1.25)). They are not equal to each other since, unlike Eqs.(29) and (30), they have no asymmetry $\bar{u} \leftrightarrow \bar{d}$. The values of these WMM elements can be found using the orthogonality of the first and third column of the obtained matrix, without having to write out the cumbersome formulae for $V_{td}$ and $V_{ub}$ as in Eqs.(29) and (30). We have:

$$V_{ts}V_{cd}^{+} \; \approx \; V_{td} + V_{ub}^{+}.\qquad(31)$$

From Eqs.(A1.24) and (A1.25), it is obvious that $V_{td}$ and $V_{ub}$ are quantities of the same order of smallness. From Eqs.(31) and (32) we therefore obtain:

$$|V_{td}| \; \sim \; |V_{ub}| \; \simeq \; \frac{m_d}{m_b},\qquad(32)$$

that is, the value much smaller than $|V_{cd}|$ and $|V_{ts}|$. The MS violation mechanism leads to the correct hierarchy in the CKM matrix. The qualitative properties of the WMM are reproduced without numerical fitting of the constants.

The abundance of free parameters in Eqs.(24) ((A1.13) – (A1.15)) and (29), (30) ($\beta^{\bar{u}}$ and $\beta^{\bar{d}}$) allows reproduction of all required numeric values of masses and $V_{ik}$. Fitting of these values is therefore always possible, but here is of no special interest.

The mechanism of mass matrix formation in the MS-model (Fig.1a) is significantly less complicated compared to the mechanism of nondiagonal-over-generations scalar vacuum averages (Fig.1b, [10]). It appears completely natural that the mass matrices $A$ and $B$ should be present in the scenario under consideration.

## 5. Lepton WMM (PMNS Matrix)

The observed neutrino mass spectrum includes two states "1,2" close to each other in mass, with the third state located far away from the first two. All neutrino masses are much lighter ($\lesssim 10^{-6}$) than the electron mass. To date, it is still uncertain whether the third neutrino is the lightest or the heaviest ("normal" or "inverse" spectrum [14,15]). With the inverse spectrum being preferable for the mirror mechanism, the neutrino spectrum is principally different from the spectrum of SM quark and charged lepton fermions.

At the same time, the hierarchy of masses appears to be a property inherent to mass spectra of all fermions. Mechanisms of its appearance do not then depend on quantum numbers or interactions (see Appendix 2). As for the light neutrino mass spectrum, its difference can be explained by the fact that the MS violation mechanism results, for these masses, in an elaborate separable equation of type Eq.(24) with its own constants $A, B, M$ (at that, $A^{(\bar{\nu})} \equiv A^{(\bar{e})}$, Eq.(4)). By fitting these constants, the modified spectrum can also be reproduced for the hierarchy of the parameters $A, B, M$, themselves being fermion masses prior to MS violation and the parameters of MS violation. It is obvious that $M(\nu)$—the masses of mirror neutrinos—must be very large.

Similarly, the lepton WMM (PMNS matrix) is absolutely different from the quark WMM (CKM matrix) [15]. In this section, we will demonstrate that the PNMS matrix can be qualitatively reproduced in the MS-model if the mass spectrum of SM neutrinos is inverse—i.e., the state "3" mass is the lightest. In this case, the PMNS matrix elements are defined only by the properties of the charged lepton system and, in the main order of the mass hierarchy for charged particles, do not depend on the parameters of the neutrino system.

Firstly, let us consider the inverse neutrino spectrum with an exact mass hierarchy ($\nu$ is omitted):

$$m_1 \gg m_2 \gg m_3 \,. \tag{33}$$

The hierarchical smallness of $m_3$ will be a necessary condition for reconciliation of lepton WMM properties. To approach the observed inverse spectrum

$$m_1 \approx m_2 \gg m_3 \tag{33'}$$

we have to consider the case of Eq.33.

In the lowest approximation of the hierarchy, the orthonormalized left-handed eigenfunctions of the separable matrix (19) depend again only on vectors $A_n$. If the hierarchy of states (33) is inverse to the masses of charged leptons, these functions for neutrino states 1-3 coincide with expressions (26) at $A_0 \leftrightarrow A_2$. We have:

$$\phi_1^{(0)} = \frac{1}{N_1} A_2 \,, \quad \phi_2^{(0)} = \frac{1}{N_2}\Big(A_1 - \frac{(A_2^+, A_1)}{|A_2|^2} A_2\Big), \quad \phi_3^{(0)} = \frac{1}{N_3}\Big[A_2^+, A_1^+\Big]. \tag{34}$$

The normal hierarchy (25) of charged leptons preserves for them the eigenfunctions (26): $III \to e_c, II \to \mu, I \to \tau$, with the same flavor-independent vectors $A$ as in Eq.(34). The scalar products (27) for functions (26) and (34) result, for leptons (33), in a WMM that is completely different from the CKM matrix. The new matrix lacks diagonal unities and the hierarchy of non-diagonal elements.

Thus, the lepton WMM problem in the mirror approach can be solved by considering the characteristics of charged leptons only, without having to consider the Yukawa or Majorana properties of neutrinos themselves: the vectors $A_n$ do not depend on the $\nu, e$-flavors.

To determine the relative positions of the $A_n$ vectors in the space of generations, it is sufficient to take the most arbitrary mass matrix of mirror analogs of charged leptons, to find matrices that diagonalize it, and to determine what happens with the diagonal matrices $\bar{A}$ , describing fermion masses prior to MS violation, as a result.

The mirror charged lepton matrices $\mu$, defined by the factors $h$ in Eq.(10), can be taken as Hermitian (parity conservation before MS violation) and correspond, per Eq.(25), to an hierarchy inverse to the spectrum of SM charged leptons.

Further, in the real lepton WMM [15], CP-violating complexes have no significant impact on the main matrix structure. Therefore, for the simplicity and clarity of interpretation, let us again consider a real symmetrical mass matrix of mirror charged leptons ($A_n$ are real three-dimensional vectors).

Such a matrix with an eigenvalue hierarchy can be built by the generalization of the known seesaw mechanism [12] for a system of three states. We have:

$$\mu = \begin{vmatrix} M & m_1 & m_2 \\ m_1 & m & 0 \\ m_2 & 0 & 0 \end{vmatrix} \sim \begin{vmatrix} M & m_2 & m_1 \\ m_2 & 0 & 0 \\ m_1 & 0 & m \end{vmatrix}, \quad M \gg m_i . \tag{35}$$

Matrix (36) is chosen in such a form that it has only one large element (energy scale). Further on, it becomes clear that this is the only necessary and important feature that guarantees the appearance of lepton WMM properties. The uniqueness of the large scale may also turn out to be a quite realistic feature of the mass hierarchy formation mechanism (see Appendix 2). By changing the generation indices in $\Psi_{LR}$ and $\Psi_{RL}$, this element can be placed in position (1,1).

The equal-to-zero elements in Eq.(35) could denote insignificant quantities that are small compared to the accountable elements. Their choice also depends on another condition of the hierarchy prompted by the seesaw mechanism—the determinant value does not contain the large scale $M$. Matrix (35) results in the characteristic equation:

$$(-\mu)^3 + (-\mu)^2(M + m) + (-\mu)(Mm - m_{12}^2) - m\,m_2^2 \;=\; 0, \quad m_{12}^2 = m_1^2 + m_2^2.$$  (36)

As known, the coefficients of this equation can be expressed through the roots $\mu$ and are respectively equal to:

$$\mu_2 + \mu_1 + \mu_0, \quad \mu_2(\mu_1 + \mu_0) + \mu_1\mu_0, \quad \mu_2\mu_1\mu_0, \quad \mu_2 \;\gg\; \mu_1 \;\gg\; \mu_0.$$  (37)

At large $M$, the roots can easily be found from Eqs.(36) and (37) with any precision. In the hierarchy of eigenvalues, the order of their magnitude is defined by the ratios of the neighboring consecutive coefficients (36). The root systems depend on the ratios between the addends in the coefficient at $(-\mu)$:

$$Mm \;>\; m_{12}^2 \quad (\text{a}), \quad Mm \;<\; m_{12}^2 \quad (\text{b}).$$  (38)

Further equations are written out for the case where inequalities (38) denote "much larger" and "much less". While not essentially changing the process of formation of the PMNS matrix properties, simplification of the formulae facilitates the understanding of the MS-scenario under consideration.

With the precision to $(m_i/M)^2$, we have for both variants in Eq.(38):

$$\begin{array}{ll}
(a): & \mu_2 = M + m_{12}^2/M \\
& \mu_1 = m - m_1^2/M \\
& \mu_0 = -m_2^2/M
\end{array}
\qquad
\begin{array}{l}
(b): \; \mu_2 = M + m_{12}^2/M \\
\;\;\;\;\; \mu_1 = -m_{12}^2/M + m(m_1^2/m_{12}^2) \\
\;\;\;\;\; \mu_0 = (m_2^2/m_{12}^2)\,m
\end{array}$$  (39)

Note that for (b), Eq. (39) results in:

$$(b): \quad m = \frac{\tilde{m}^2}{M} \;<\; \frac{m_{12}^2}{M},$$  (40)

so that in our formulae for variant (b), $m \ll (m_{12}^2/M)$. As in the seesaw mechanism, the sign of the mass is irrelevant in the given scenario. The difference between the two variants, (a) and (b), becomes evident at $M \to \infty$. In (a), besides $M \to \infty$, there remains one more finite scale $m$ in the formulae. In (b), all masses, except $M \to \infty$, vanish. This makes variant (b) entirely consistent with the seesaw mechanism.

By finding eigenfunctons for each of the roots in (a) and (b), let us construct the orthogonal matrices $U$ (at the same precision of $(m_i/M)^2$), diagonalizing Eq.(35):

$$(a): \quad U = \begin{vmatrix} 1 & \dfrac{m_1}{M} & \dfrac{m_2}{M} \\ \dfrac{m_1}{M} & -1 & -\dfrac{m_1 m_2}{mM} \\ \dfrac{m_2}{M} & \dfrac{m_1 m_2}{mM} & -1 \end{vmatrix}; \quad (b): \quad U = \begin{vmatrix} 1 & \dfrac{m_{12}}{M} & \dfrac{mm_1 m_2}{m_{12}^3} \\ \dfrac{m_2}{M} & -\dfrac{m_2}{m_{12}} & \dfrac{m_1^2}{m_{12}^2} \\ \dfrac{m_1}{M} & -\dfrac{m_1^2}{m_{12}} & -\dfrac{m_2}{m_{12}} \end{vmatrix}. \qquad (41)$$

Let us take a diagonal form of the isodoublet mass matrix: $\tilde{A} = \left(\tilde{A}_1, \tilde{A}_2, \tilde{A}_3\right)$ in Eq.(3) and transform it with the matrix $U$, as $U\tilde{A}U^+$. For the vector-columns $A_2, A_1, A_0$ in Eq.(19), we have for the lepton mass matrices $\ell$ and $v$:

(a):

$$A = \begin{vmatrix} & A_2 & A_1 & A_0 \\ \hline & \tilde{A}_1 & \left(\tilde{A}_1 - \tilde{A}_2\right)m_1/M & \left(\tilde{A}_1 - \tilde{A}_3\right)m_2/M \\ & \left(\tilde{A}_1 - \tilde{A}_2\right)m_1/M & \tilde{A}_2 & \left(\tilde{A}_3 - \tilde{A}_2\right)\dfrac{m_1 m_2}{mM} \\ & \left(\tilde{A}_1 - \tilde{A}_3\right)m_2/M & \left(\tilde{A}_3 - \tilde{A}_2\right)\dfrac{m_1 m_2}{mM} & \tilde{A}_3 \end{vmatrix}; \qquad (42)$$

(b)

$$A = \begin{vmatrix} & A_2 & A_1 & A_0 \\ \hline & \tilde{A}_1 & \left(\tilde{A}_1 - \tilde{A}_2\right)\dfrac{m_2}{M} + \tilde{A}_3 \dfrac{m_1^2 m_2 m}{m_{12}^3} & \left(\tilde{A}_1 - \tilde{A}_2\right)\dfrac{m_1}{M} - \tilde{A}_3 \dfrac{m_2^2 m_1 m}{m_{12}^3} \\[2mm] & \left(\tilde{A}_1 - \tilde{A}_2\right)\dfrac{m_2}{M} + \tilde{A}_3 \dfrac{m_1^2 m_2 m}{m_{12}^3} & \tilde{A}_3 \dfrac{m_1^2}{m_{12}^2} + \tilde{A}_2 \dfrac{m_2^2}{m_{12}^2} & \left(\tilde{A}_2 - \tilde{A}_3\right)\dfrac{m_1 m_2}{m_{12}^2} \\[2mm] & \left(\tilde{A}_1 - \tilde{A}_2\right)\dfrac{m_1}{M} - \tilde{A}_3 \dfrac{m_2^2 m_1 m}{m_{12}^3} & \left(\tilde{A}_2 - \tilde{A}_3\right)\dfrac{m_1 m_2}{m_{12}^2} & \tilde{A}_3 \dfrac{m_2^2}{m_{12}^2} + \tilde{A}_2 \dfrac{m_1^2}{m_{12}^2} \end{vmatrix} \qquad (43)$$

In the lowest approximation of the mass hierarchy in (a), Eq.(42) leads to a diagonal matrix, i.e., to vectors $A_n$ that are orthogonal in the space of generation indices. For any hierarchy of neutrino masses, the WMM, Eq.(27), proves to be a unit diagonal matrix, which is inappropriate for the formation of the PMNS matrix.[4]

---

[4] It is possible that variant (a) can be related with the case of normal hierarchy of both flavors $\bar{u}$ and $\bar{d}$, i.e., with the formation of the *CKM* matrix for quarks.

For Eq. (43) (b), the matrix we obtain in the lowest approximation of the hierarchy is:

$$
A \;=\;
\begin{array}{|ccc|}
A_2 & A_1 & A_0 \\
& & \\
\tilde{A}_1 & 0 & 0 \\
& & \\
0 & \tilde{A}_3 \frac{m_1^2}{m_{12}^2} + \tilde{A}_2 \frac{m_2^2}{m_{12}^2} & \left(\tilde{A}_2 - \tilde{A}_3\right)\frac{m_1 m_2}{m_{12}^2} \\
& & \\
0 & \left(\tilde{A}_2 - \tilde{A}_3\right)\frac{m_1 m_2}{m_{12}^2} & \tilde{A}_3 \frac{m_2^2}{m_{12}^2} + \tilde{A}_2 \frac{m_1^2}{m_{12}^2}
\end{array}
\tag{44}
$$

Fig.2 depicts vectors $A_n$ from Eq.(44) and the orthonormalized vectors (26) and (34), which, in the approximation under consideration, are wavefunctions of SM particles. Mind the correspondence of directions (for the different mass hierarchies of $\nu$ and $\ell$):

$$
\begin{aligned}
&\text{For } \nu: && \nu_1 \sim A_2\,,\ \nu_2 \sim A_1 - \cos\alpha_{12} A_2\,,\ \nu_3 \sim [A_1, A_2]; \\
&\text{For } \ell^{\pm}: && \tau \sim A_0\,,\ \mu \sim A_1 - \cos\alpha_{01} A_0\,,\ e \sim [A_0, A_1]\,;
\end{aligned}
\tag{45}
$$

where $\alpha_{ik}$ are angles between $A_i$ and $A_k$. According to Eq.(44), the vector $A_2$ is orthogonal to the vectors $A_0$ and $A_1$, $\cos\alpha_{12} = 0$. In Fig.2, $A_2$ is directed along the Z axis.

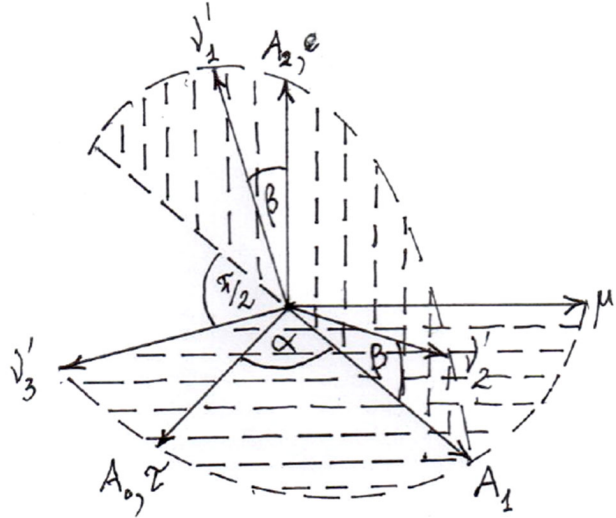

Fig.2: Formation of the carcass of the PMNS lepton WMM as rotation between axes, Eq.(46). In the space of generations: $A_2 \perp A_0, A_1$; angles $\alpha, \beta$ are determined in Eqs. (47),(48).

The mixing matrix is the matrix of the transition from the basis $\nu$ to basis $\ell$, Eq.(45). In the approximation being considered, we have:

$$V = \begin{vmatrix} 1 & 0 & 0 \\ 0 & \cos\alpha_{01} & \sin\alpha_{01} \\ 0 & -\sin\alpha_{01} & \cos\alpha_{01} \end{vmatrix}. \qquad (46)$$

Eq.(46) would serve as the basis for the lepton WMM if the neutrino spectrum obeyed a strictly hierarchical order (33). The subsequent higher order approximations would lead to small values for elements equal to zero in Eq.(46) and insignificant changes in the other elements.

For the real spectrum (33'), it is necessary to perform an additional rotation of the coordinates in Fig.2 even for building the "skeleton" of the WMM in the lowest approximation. In fact, we can consider the situation (33') as a degeneracy of levels "1,2" and solve the problem of degeneracy removal. This problem was studied in the first paper [13], which established conditions imposed on the parameters $A$ and $B$ that helped achieve the mass equality $m_1 \sim m_2$ in the zero approximation, and identified corrections removing degeneracy. The next step of the problem consists in determining the correct wavefunctions $\phi_1'$ and $\phi_2'$, once again in the zero approximation. The result is well known; it is reduced to rotation of the degeneracy problem functions $\phi_1^0$ and $\phi_2^0$ by a certain angle $\beta_{12}$ in their plane (Fig.2). In our case, these functions coincide with (34); as well, they do not depend on flavor for spectrum (33'). We have:

$$\begin{aligned} \phi_1' &= \phi_1^{(0)} \cos\beta_{12} + \phi_2^{(0)} \sin\beta_{12}, \\ \phi_2' &= -\phi_1^{(0)} \sin\beta_{12} + \phi_2^{(0)} \cos\beta_{12}. \end{aligned} \qquad (47)$$

But now the angle $\beta_{12}$ is a function of the parameters $A$ and $B^\nu$. Therefore, it depends on the flavor $\nu$. However, the actual type of such dependence does not modify the general structure of the matrix $V$ and, as such, is of no consequence. After rotation (47), the "skeleton" of the lepton WMM assumes the form:

$$V = \begin{vmatrix} 1 & 2 & 3 \\ \cos\beta & \sin\beta & 0 \\ -\sin\beta\cos\alpha & \cos\beta\cos\alpha & \sin\alpha \\ \sin\beta\sin\alpha & -\cos\beta\sin\alpha & \cos\alpha \end{vmatrix} \begin{matrix} e \\ \mu \\ \tau \end{matrix} \qquad (48)$$

In (48) we omitted the indices of the angles $\alpha, \beta$. From (48) it is clear that the term $V_{e3} = |sin\theta_{13}|$ [15] is non-vanishing only if we also take into consideration the small terms of the subsequent

mass hierarchy orders. To estimate its value, we must consider not only the terms of the known ratios of charged lepton masses [15]:

$$\frac{m_\ell}{m_\mu} \simeq \frac{1}{207} \simeq 0.0018\,, \quad \frac{m_\mu}{m_\tau} \simeq \frac{1}{17} \simeq 0.056\,, \tag{49}$$

but also corrections for the neutrino wavefunctions (34) and (47) induced, in the inverse variant, by the tiny, unknown neutrino mass $m_3 \ll m_1, m_2$. Too small corrections from charged leptons in (49) may be insufficient for the description of the observed $|sin\theta_{13}| \sim 0.14 - 0.16$. The masses $m_1, m_2$ can be calculated here from the observed $\Delta m_{ik}^2$. Then:

$$m_3 \approx 0.01 \ \text{эВ}\,, \quad \frac{m_3}{m_{1,2}} \lesssim 0.2\,, \quad m_1 \sim m_2 \sim 0.05 \ \text{эВ}\,, \tag{50}$$

will exactly suit our purpose. The remaining terms of $V$ are easily matched with the respective values of the PMNS matrix terms. Small corrections will have a very minor effect on the "skeleton" (48).

The small value of $m_3 \ll m_1, m_2$ is a necessary condition for the appearance of structure (48). The abandonment of this smallness would require including in the wavefunctions (34) and (47) terms that are not exclusively determined by the factors $A$, independent of the $(\nu, \text{l})$ flavor. Experimental knowledge of only neutrino mass squared differences leaves such a possibility open. This, however, would also mean a complete modification of the resultant structure of the matrix $V$.

## 6. Conclusion

Rather than developing the dynamics of the chosen system (3), (7), the objective of the model under consideration is to pick favorable conditions for the appearance of fermion mass matrices that can naturally reproduce both the observed structure of mass spectra and the properties of the corresponding WMM. The MS-approach is most commonly utilized to resolve the paradox with a possible use of physical methods for left/right determination in the SM. An easy simultaneous reproduction of WMM properties seems here to be very attractive. It is, obviously, difficult to imagine dynamical mechanisms responsible for the appearance of the lepton WMM's inconspicuous characteristics. The spontaneously violated MS scenario can provide a plausible path in this direction.

It is clear that further discussions are required to develop a real dynamical scheme. The proposed mechanism of the MS-violation—the direct copy of the SM—cannot be considered as absolutely appropriate. Certainly, very large masses of mirror analogs of SM particles would appear to be able to significantly diminish the influence of mirror particles on SM processes. At the same time, the real Higgs boson $H$ in the MS violation scenario under consideration is non-perturbatively related with mirror fermions: $h = M/\eta \gg 1$, $M \gg \eta = 246$ GeV, which leads to an exclusively strong coupling. However, the same strong interaction of $H$ gives hope that their role in the appearance of the Higgs boson (through loops of virtual mirror fermions) will be significantly diminished [18]. At that, transitions of light SM particles into mirror states (parameters $A, B$) could remain perturbative.

Yet, the MS violation mechanism proposed in this paper can help perceive the possibilities (in addition to the main purpose) inherent in the MS.

Diagonalization of mass terms (19) for light particles does not generally affect the diagonality over generations of electroweak and other flavor-neutral interactions. This is easily confirmed by the flavor independence of the left functions (26) in the lowest orders of the mass hierarchy and by the unitarity of transformation matrices.

The Higgs boson $\Phi_1(H)$ interacts directly only with the heavy components $\Psi$, while interacting with the light $\psi$ only through small wavefunction terms (15). So, diagonalization of the Yukawa constants (10) and matrices (19) also leads, in the lowest approximation of the mass hierarchy, to the diagonal interaction of light SM-states with the Higgs particle $H$. Non-diagonal terms, as well as right-handed currents, can appear here only in subsequent orders. The second Higgs boson $\Phi_2$ should be assumed to be very heavy (the choice of the constant $k$ in Eq.(11)).

Non-diagonal-over-generations couplings of mirror particles with SM-fermions, which appear in subsequent orders of the mass hierarchy, make mirror states unstable. At present, the character of neutrinos, Dirac or Majorana, and the type of their spectrum, normal or inverse, have not yet been established with certainty [14,15]. Our approach to the lepton WMM leans in favor of the Dirac character and inverse spectrum of neutrinos. For Majorana neutrinos, the proposed mechanism of PNMS matrix formation is thought to be unrealizable as it requires introducing additional, unsubstantiated conditions on the model constants (such as the simultaneous mutual diagonalization of the Majorana and Dirac parts of mirror neutrino masses, the equality of the matrices $A = B^{(\nu)}$, etc.).

# Appendix 1

*A. Mass spectrum calculation*

Let us simplify the expression for the sum of separable matrices (19), including the value $1/M_n^{(f)}$ in the parameter $B_n^{(f)}$. The form of the mass matrix (19) used here is:

$$(\mathcal{M}_{LR})_a^b = \sum_{n=0}^{2} A_a^n B_n^{+b}. \tag{A1.1}$$

The arbitrary matrix $\mathcal{M}$ is used to build the Hermitian matrix:

$$(\mathcal{M}\mathcal{M}^+)_a^{a'} = (\mathcal{M}_{LR})_a^b (\mathcal{M}_{LR}^+)_b^{a'} = \sum_{n,n'} A_a^n B_n^{+b} B_b^{n'} A_{n'}^{+a'}. \tag{A1.2}$$

The separable form results here in (A1.2) being a product of matrices built on projections of the vectors $A_n$, $B_n$, $n = 0,1,2$, in the space of generation indices $a, b = 1,2,3$:

$$(\mathcal{M}\mathcal{M}^+)_{LL} = \Delta_3(A_0, A_1, A_2)\Delta_3(B_0^+, B_1^+, B_2^+)\Delta_3(B_0, B_1, B_2)\Delta_3(A_0^+, A_1^+, A_2^+); \tag{A1.3}$$

where the matrix $\Delta_3$ equals

$$\Delta_3(A_0, A_1, A_2) = \begin{vmatrix} A_1^0 & A_1^1 & A_1^2 \\ A_2^0 & A_2^1 & A_2^2 \\ A_3^0 & A_1^1 & A_1^2 \end{vmatrix}, \Delta_3(A_0^+, A_1^+, A_2^+) = \Delta_3^+(A_0, A_1, A_2). \tag{A1.4}$$

Any minors and the determinant of the separable matrix, e.g., (A1.1), are again the product of determinants of the respective order. This follows from the formula for the arbitrary minor of the $N^{\text{th}}$ order (with an arbitrary number of indices):

$$\varepsilon_{aa'a''}...\mathcal{M}_a^b \mathcal{M}_{a'}^{b'}...\frac{1}{N!}\varepsilon_{a_1 a_2 a_3}...\mathcal{M}_b^{+a_1}\mathcal{M}_{b'}^{+a_2}\mathcal{M}_{b''}^{+a_3} \dots. \tag{A1.5}$$

Since all $a, a', a'', ...$ must be different, then $b, b', b'', ...$ (as well as the indices $n, n'$ in (A1.2) are also different. With $\mathcal{M}$ having a separable structure, expressions (A1.2) and their determinants are products of matrices $\Delta_N$ and, consequently, respective determinants $D_N$. In $\mathcal{M}$ and $\mathcal{M}^+$, the indices $b$ are the same throughout; summation is performed over these indices, therefore the result of calculations will contain matrices and determinants of the type:

$$\Delta_3(B_0, B_1, B_2)\Delta_3(B_0^+, B_1^+, B_2^+) \;=\; \Delta_3\Big(|B_0|^2, |B_1|^2, |B_2|^2\Big), \tag{A1.6}$$

where the matrix $\Delta_3(|B_0|^2, |B_1|^2, |B_2|^2)$ is based on scalar products of the vectors $(B_i^+, B_k)$:

$$\Delta_3(|B_0|^2, |B_1|^2, |B_2|^2) \;=\; \begin{vmatrix} |B_0|^2 & (B_0^+, B_1) & (B_0^+, B_2) \\ (B_1^+, B_0) & |B_1|^2 & (B_1^+, B_2) \\ (B_2^+, B_0) & (B_2^+, B_1) & |B_2|^2 \end{vmatrix} \tag{A1.7}$$

Only diagonal elements are written out in the matrix arguments. The same principle is used in other matrices and determinants built at different $N$.

After that, we can easily write out the coefficients of the characteristic equation of the matrix (A1.1), $\rho = m^2$, as follows:

$$(-\rho)^3 + \rho^2(\mathrm{Tr}\,\mathcal{M}\mathcal{M}^+) - \rho \;\;(\text{sum of main minors}) \;+ \det \mathcal{M}\mathcal{M}^+ = 0, \tag{A1.8}$$

where the trace of the matrix equals (with contributions arranged by the degree of smallness in accordance with (25) ($|B_0| \gg |B_1| \gg |B_2|$)):

$$\mathrm{Tr}\,\mathcal{M}\mathcal{M}^+ = |A_0|^2|B_0|^2 + 2\mathrm{Re}[(A_1^+, A_0)(B_0^+, B_1)] + |A_1|^2|B_1|^2 +$$
$$+ 2\mathrm{Re}[(A_2^+, A_0)(B_0^+, B_2)] + 2\mathrm{Re}[(A_2^+, A_1)(B_1^+, B_2)] + |A_2|^2|B_2|^2. \tag{A1.9}$$

The sum of the main minors is equal to:

$$\sum \;=\; D_2(|A_0|^2, |A_1|^2)D_2(|B_0|^2, |B_1|^2) + 2\mathrm{Re}\left[D_2(|A_0|^2, (A_1^+, A_2))D_2(|B_0|^2, (B_2^+, B_1))\right] +$$
$$+ 2\mathrm{Re}\left[D_2\left((A_0^+, A_1), (A_1^+, A_2)\right)D_2\left((B_1^+, B_0), (B_2^+, B_1)\right)\right] + D_2(|A_0|^2, |A_2|^2)D_2(|B_0|^2, |B_2|^2) +$$
$$+ 2\mathrm{Re}\left[D_2\left((A_0^+, A_1), |A_2|^2\right)D_2\left((B_1^+, B_0), |B_2|^2\right)\right] + D_2(|A_1|^2, |A_2|^2)D_2(|B_1|^2, |B_2|^2). \tag{A1.10}$$

In Eq.(A1.10), the determinant $D_2$ from the scalar products of four vectors is calculated as follows:

$$D_2\Big((a, b), (c, d)\Big) = \begin{vmatrix} (a, b) & (c, b) \\ (a, d) & (c, d) \end{vmatrix}. \tag{A1.11}$$

Finally, the determinant $\mathcal{M}\mathcal{M}^+$ equals (see (A1.2)

$$\det \mathcal{M}\mathcal{M}^+ \;=\; D_3\Big(|A_0|^2, |A_1|^2, |A_2|^2\Big) D_3\Big(|B_0|^2, |B_1|^2, |B_2|^2\Big). \tag{A1.12}$$

It is obvious that Eq. (A1.12) is symmetrical: $R \leftrightarrow L$.

To confirm the WMM properties outlined earlier in this paper, it is necessary to calculate the values $\rho_I$, $\rho_{II}$ and $\rho_{III}$ at various precisions. For the larger mass $\rho_I$, we have to calculate three successive perturbation theory orders (25), while for the smaller masses $\rho_{II}$ and $\rho_{III}$, calculation of two orders is sufficient. As a result, we obtain:

$$
\begin{aligned}
m_I^2 \;=\; & |A_0|^2 |B_0|^2 + 2\mathrm{Re}\Big[(A_0^+, A_1)(B_0, B_1^+)\Big] + 2\mathrm{Re}\Big[(A_0^+, A_2)(B_0, B_2^+)\Big] + \\
& + |A_1|^2 |B_1|^2 - \frac{D_2(|A_0|^2, |A_1|^2) D_2(|B_0|^2 |B_1|^2)}{|A_0|^2 |B_0|^2} \;+\; \cdots
\end{aligned}
\tag{A1.13}
$$

$$
\begin{aligned}
m_{II}^2 \;=\; & \frac{D_2(|A_0|^2, |A_1|^2) D_2 |B_0|^2, |B_1|^2)}{|A_0|^2 |B_0|^2}\Big\{ 1 - \frac{2\mathrm{Re}[(A_0^+, A_1)(B_1^+, B_0)]}{|A_0|^2 |B_0|^2} + \\
& + \frac{2\mathrm{Re}[D_2(|A_0|^2, (A_1^+, A_2)) D_2(|B_0|^2, (B_2^+, B_1))]}{D_2(|A_0|^2, [A_1|^2) D_2(|B_0|^2, |B_1|^2)} + \cdots \Big\}.
\end{aligned}
\tag{A1.14}
$$

$$
\begin{aligned}
m_{III}^2 \;=\; & \frac{D_3(|A_0|^2, |A_1|^2, |A_2|^2) D_3(|B_0|^2, |B_1|^2, |B_2|^2)}{D_2(|A_0|^2, |A_1|^2) D_2(|B_0|^2, |B_1|^2)} \times \\
& \times \Big\{ 1 - \frac{2\mathrm{Re}[(D_2(|A_0|^2, (A_1^+, A_2)) D_2(|B_0|^2, (B_2^+, B_1))]}{D_2(|A_0|^2, |A_1|^2) D_2(|B_0|^2, |B_1|^2)} + \cdots \Big\}.
\end{aligned}
\tag{A1.15}
$$

The expressions beforethe brace in (A1.14) and (A1.15) represent the lowest, non-zero approximations for each of the masses $II$ and $III$. The first term in (A1.13) is an approximate square of the larger mass $I$. All mass squares are greater than zero. Formulae for their lowest approximations are written out in (24).

### B. Eigenfunctions of matrix (A1.2)

To calculate the WMM (Eq.(27):

$$
V_{TS} \;=\; \Big(\phi_{LT}^{(\bar{u})}, \phi_{LS}^{(d)}\Big), \quad T, S \;=\; I, II, III,
\tag{A1.16}
$$

where $T, S$ are the numbers of physical generations, $I$ is the heaviest, i.e., the third in normal ordering: $(tb)$, $(cs)$, $(ud)$. It is necessary to find the left eigenfunctions $\phi_l$ of operators (19) or (A1.1). This is a much more cumbersome task than that of the spectrum, since the functions are vectors in the space of the indices $a, b$. Not only do they depend on scalar products but also directly on vector components.

Diagonalizing matrix M equations with a determinant equal to zero define the ratio of vector components in the solutions (here $\mathcal{M} = \mathcal{M}\mathcal{M}^+$):

$$\frac{x_1}{x_3} = \frac{\rho\mathcal{M}_{13} - (\mathcal{M}_{13}\mathcal{M}_{22} - \mathcal{M}_{12}\mathcal{M}_{23})}{\rho^2 - \rho(\mathcal{M}_{11} - \mathcal{M}_{22}) + (\mathcal{M}_{11}\mathcal{M}_{22} - \mathcal{M}_{21}\mathcal{M}_{12})};$$ (A1.17)

and similarly for $x_2$, $1 \leftrightarrow 2$. To isolate the dependence on indices in the numerator and denominator (A1.17) in all three perturbation orders at once, one has to write out each component $A_a^n$ of the matrix $\mathcal{M}$, and these contain many terms. For example,

$$\begin{aligned}
\mathcal{M}_1^3 = & \ A_1^0\Big[(B_0^+, B_0)A_0^{+3} + (B_0^+, B_1)A_1^{+3} + (B_0^+, B_2)A_2^{+3}\Big] + \\
& + A_1^1\Big[(B_1^+, B_0)A_0^{+3} + (B_1^+, B_1)A_1^{+3} + (B_1^+, B_2)A_2^{+3}\Big] + \\
& + A_1^2\Big[(B_2^+, B_0)A_0^{+1} + (B_2^+, B_1)A_1^{+3} + (B_2^+, B_2)A_2^{+3}\Big].
\end{aligned}$$ (A1.18)

In the wavefunctions $\phi_{LS}$, it is necessary to find three terms of the hierarchy for $S = I$ and two terms, for each of the $II$ and $III$ states. Calculations result in the following, comparatively simple formulae for the state vectors:

$$\phi_I = \frac{1}{|A_0|}\left\{A_0 + \frac{(B_1^+, B_0)}{|B_0|^2}\left(1 + \frac{2\text{Re}((A_0, A_1^+)(B_0^+, B_1))}{|A_0|^2|B_0|^2}\right)A_1 + \frac{(B_2^+, B_0)}{|B_0|^2}A_2\right\},$$ (A1.19)

$$\phi_{II} = \frac{|A_0|}{|(D_2(|A_0|^2, |A_1|^2)|^{1/2}}\left\{A_1 - \frac{(A_0^+, A_1)}{|A_0|^2}A_0 - \frac{D_2(|A_0|^2, |A_1|^2)}{|A_0|^4}\frac{(B_0^+, B_1)}{|B_0|^2}A_0 - \right. \\ \left. - \frac{D_2(|B_0|^2, (B_2^+, B_1))}{D_2(|B_0|^2, |B_1|^2)}\frac{[A_0^+[A_0, A_2]]}{|A_0|^2}\right\},$$ (A1.20)

$$\phi_{III} = \frac{1}{|D_2(|A_0|^2, |A_1|^2)|^{\frac{1}{2}}}\left\{[A_0^+, A_1^+] + \frac{D_2(|B_0|^2, (B_1^+, B_2))}{D_2(|B_0|^2, |B_1|^2)}[A_0^+, A_2^+] + \right. \\ \left. + \frac{D_2((B_0^+, B_1)(B_1^+, B_2))}{D_2(|B_0|^2, |B_1|^2)}[A_1^+, A_2^+]\right\}.$$ (A1.21)

The square brackets designate vector products.

The normalizing multipliers in these formulae are written out in the lowest order approximation, which is sufficient for our purposes. Therefore, in the lowest, other-than-zero order, the multipliers and all $\phi_T$ depend only on vectors $A_n$. Under condition (4) $A^{\bar{u}} \equiv A^{\bar{d}}$, their orthonormality results in the diagonal elements $V_{TT} \simeq 1$. Non-diagonal elements belong to subsequent orders of the hierarchy; they are defined not by the main terms in $\phi_T$ but rather by corrections depending on $B^{\bar{u}} \neq B^{\bar{d}}$. Thus, the hierarchy of the mass matrix gives rise to the WMM hierarchy. It is clear that condition (4) plays a key role in this scenario.

## C. Weak mixing matrix

Eqs.(A1.19)-(A1.21) allow calculation of the scalar products (A1.16) constituting the CKM matrix. As we remember, owing to (4), $\bar{u}$ and $\bar{d}$ are different only in the right factors $B_n$.

For quantities included in the WMM formula we have:

$$V_{ts} = \left(\phi_I^{(\bar{u})+}, \phi_{II}^{(d)}\right) = \frac{D_2((|A_0|^2, |A_1|^2))^{\frac{1}{2}}}{|A_0|^2} \left\{ \left(\frac{(B_0^+, B_1)}{|B_0|^2}\right)_{\bar{u}} - \left(\frac{(B_0^+, B_1)}{|B_0|^2}\right)_{\bar{d}} \right\}; \tag{A1.22}$$

$$V_{cd}^+ = \left(\phi_{II}^{(\bar{u})+}, \phi_{III}^{(d)}\right) = \frac{|A_0|(A_0^+, [A_1^+, A_2^+])}{D_2(|A_0|^2, |A_1|^2)} \times$$
$$\times \left\{ \left(\frac{D_2(|B_0|^2, (B_1^+, B_2))}{D_2(|B_0|^2, |B_1|^2)}\right)_{\bar{u}} - \left(\frac{D_2(|B_0|^2, (B_1^+, B_2))}{D_2(|B_0|^2, |B_1|^2)}\right)_{\bar{d}} \right\}. \tag{A1.23}$$

The coefficients $V_{ts}$ and $V_{cd}$ are first-order quantities. The coefficient of the following order of smallness $V_{td}$ equals:

$$V_{td} = \left(\phi_I^{(\bar{u})+}, \phi_{III}^{(d)}\right) = \frac{(A_0^+, [A_1^+, A_2^+])}{|A_0|(D_2(|A_0|^2, |A_1|^2))^{1/2}} \left\{ \left(\frac{(B_0^+, B_2)}{|B_0|^2}\right)_{\bar{u}} + \right.$$
$$\left. + \left(\frac{D_2((B_0^+, B_1), (B_1^+, B_2))}{D_2(|B_0|^2, |B_1|^2)}\right)_{\bar{d}} - \left(\frac{(B_0^+, B_1)}{|B_0|^2}\right)_{\bar{u}} \left(\frac{D_2(|B_0|^2, (B_1^+, B_2))}{D_2(|B_0|^2, |B_1|^2)}\right)_{\bar{d}} \right\}. \tag{A1.24}$$

The elements $V_{ub}^+ = V_{td}(\bar{u} \leftrightarrow \bar{d})$ differ from (A1.24) because there is no symmetry (anti) $\bar{u} \leftrightarrow \bar{d}$ here:

$$V_{ub}^+ = \frac{(A_0^+, [A_1^+, A_2^+])}{|A_0|(D_2(|A_0|^2, |A_1|^2))^{1/2}} \left\{ \left(\frac{(B_0^+, B_2)}{|B_0|^2}\right)_{\bar{d}} + \left(\frac{D_2((B_0^+, B_1), (B_1^+, B_2))}{D_1(|B_0|^2, |B_1|^2)}\right)_{\bar{u}} - \right.$$
$$\left. - \left(\frac{(B_0^+, B_1)}{|B_0|^2}\right)_{\bar{d}} \left(\frac{D_2(|B_0|^2, (B_1^+, B_2))}{D_2(|B_0|^2, |B_1|^2)}\right)_{\bar{u}} \right\}. \tag{A1.25}$$

Eq.(A1.25) is written out to facilitate demonstration of the orthogonality of the resultant matrix $V$ columns, that is, Eq.(32).

For the sum $V_{td} + V_{ub}^+$, we group the terms relating only to the flavor $\bar{u}$ and those relating only to the flavor $\bar{d}$, and write out the determinants:

$$D_2\Big((B_0^+, B_1), (B_1^+, B_2)\Big) = (B_0^+, B_1)(B_1^+, B_2) - |B_1|^2, (B_0^+, B_2),$$

$$D_2\Big(|B_0|^2, |B_1|^2\Big) = |B_0|^2|B_1|^2 - (B_0^+, B_1)(B_1^+, B_0).$$

(A1.26)

We obtain the product of expressions (A1.22) and (A1.23), i.e., Eq.(31). Of note, neither phases nor absolute values of the parameters were fitted.

For the phases, we had to select the function phases $\phi^{(\bar{u})}$ and $\phi^{(\bar{d})}$ in Eqs.(A1.19)-(A1.21) so that the diagonal elements in the lowest order $(\phi^{(\bar{u})^+}, \phi^{(\bar{d})^+})$ were equal to unity. It would have been illogical at this point, albeit with no implications for the physics, to break the $SU(2)$-symmetry. For Eq.(31), we could also have left these phases arbitrary. The total number of remaining parameters in the mass matrix representation (19) permits reproduction of any WMM, including CP violation phases. The correct hierarchy of CKM matrix elements automatically arises from the quark mass hierarchy and consequences of the weak $SU(2)$-symmetry, without a need for direct numerical fitting of the parameters.

## Appendix 2

The hypothesis that the appearance of fermion mass spectrum hierarchy has some universal cause independent of the nature of interactions is supported by one strong phenomenological evidence: all observed SM fermion generations (perhaps with the only exception of neutrinos) have hierarchical mass spectra.

Theoretical physicists have discussed mass hierarchy for many decades. The first, and best known, work in this field was by Froggatt and Nielsen [21], which has since been advanced by them [10] and many others [22]. An extensive bibliography can be found in a more recent paper [23]. The focus of all work in this area has been to devise interactions and fit their properties such that, upon violation of the chiral symmetry of a massless system, it was possible to reproduce the observed mass hierarchies.

It is important to note here that there is a possible relationship between the character of the fermion mass spectrum and the depth of the system's Dirac sea. When $n$-fermions acquire masses, the vacuum energy is reduced by

$$-\sum_{\mathbf{p},n}\Big(\sqrt{\mathbf{p}_n^2 + m_n^2} - |\bar{p}_n|\Big) \simeq -\sum_{\mathbf{p}} \frac{\sum_n m_n^2}{|\mathbf{p}|}.$$

(A2.1)

The scale of chiral symmetry violation in a system of $n$-fermions is dependent on $\mathrm{Tr}\mathcal{M}$, where $\mathcal{M}$ is the mass matrix of fermions. In the hypothetical characteristic equation for masses, in the case of the hierarchical spectrum, its other invariants will not represent the general scale of the phenomenon.

The sum of mass squares

$$\sum_n m_n^2 \;=\; (\mathrm{Tr}\mathcal{M})^2 - \sum_{n \neq n'} 2m_n m_{n'}\,. \tag{A2.2}$$

is maximum at a very large mass of only one of the particles:

$$m_1 \simeq \mathrm{Tr}\,\mathcal{M}, \quad m_2, m_3 \ldots \ll \; m_1\,. \tag{A2.3}$$

Certainly, Eq.(A2.1) uses energies of free particles; the sums there diverge. However, the speculation has relevance to the precise Dirac vacuum of the system – the divergence means a principal contribution from very high energy particles, which are not much affected by interactions.

It is clear that establishing the relationship between the vacuum depth and the character of particle spectrum may present a very difficult problem, practically with no solution. Moreover, Eqs.(A2.1)-(A2.3) should be supplemented with conditions to prevent other masses to vanish. The hierarchical character of relationships between these masses can also be due to energy reasons.

It, however, appears important that the resulting relationship between vacuum and spectrum could be independent of the properties of interactions.